\documentclass{ws-procs9x6}
\begin{document}

\title{AT WHICH ORDER SHOULD WE TRUNCATE PERTURBATIVE SERIES?}

\author{Y. MEURICE}

\address{Department of Physics and Astronomy, University of Iowa,\\
Iowa City, IA 52242, USA\\
$^*$E-mail: yannick-meurice@uiowa.edu\\}

\begin{abstract}
Perturbative coefficients grow factorially with the order and one needs a prescription to truncate the series in order to obtain a finite result. A common prescription 
consists in dropping the smallest contribution at a given coupling and all the higher orders  terms. We discuss the error associated with this procedure. 
We advocate a more systematic approach which consists in controlling the large field 
configurations in the functional integral. We summarize our best understanding of these issues for lattice QCD in the quenched approximation and their connection with convergence problems found in the continuum.

\end{abstract}


\bodymatter

\section{Introduction}

Perturbation theory has played an essential role in developing and establishing the 
standard model of electroweak and strong interactions. The renormalizability of the 
theory guarantees that we can calculate the radiative corrections at any order in perturbation theory. On the other hand, a generalization of Dyson's argument \cite{dyson52}
suggests that the perturbative series are divergent and one needs to truncate the series. In absence of a definite prescription to deal with this problem, one usually relies on the ``rule of thumb'' which consists in dropping the smallest 
contribution at a given coupling and all the higher order terms. Clearly, this procedure has a limited accuracy and it 
is not always obvious how to estimate the error or to decide if one needs to calculate one extra order. 

The problem is particularly acute for QCD corrections because they are large even at low order. As emphasized by Z. Bern's talk \cite{zvibern}, NLO corrections are important for 
multijet processes to be studied by the LHC. Another example \cite{larin94} is 
the hadronic width of the $Z^0$ where 
the term of order $\alpha_s^3$ 
is more than 60 percent of the 
term of order $\alpha_s^2$ and contributes to one part in 1,000 
of the total width (a typical experimental error for individual LEP experiments). It is not 
clear that the next term would improve the accuracy of the calculation. As these calculations 
can be extremely time consuming, it is necessary to address the lack of convergence of 
perturbative series in a more systematic way.

\section{The Rule of Thumb}
\begin{figure}
\includegraphics[width=0.7\textwidth]{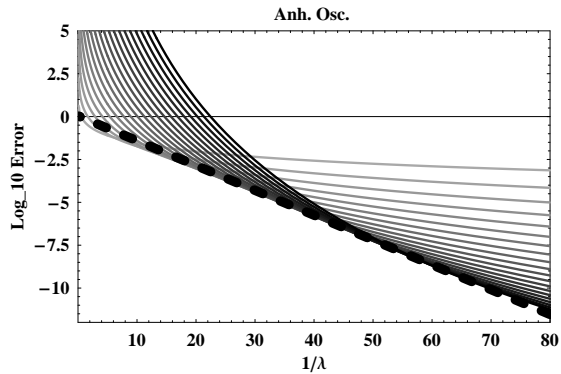}
\caption{Absolute value of the difference between the 
series and the numerical value for order 1 to 15 (in a $Log_{10}$ scale) for the anharmonic oscillator as a function of $1/\lambda$. As the order increases, 
the curves get darker. The dash curve is  (${\rm log}_{10}(\sqrt{12}/\pi){\rm e}^{-\frac{1}{3\lambda}}$) }
\label{fig:anh}
\end{figure}

Consider a generic asymptotic series in a coupling $\lambda$ with coefficients growing like $ C_1C_2^k\Gamma(k+C_3)$. If we assume that for sufficiently small $\lambda$ the error made by
truncating the series at order $k$ is given by the order $k+1$ contribution, it is 
possible to show that the error is minimized by truncating the series at order $(\lambda |C_2|)^{-1}-C_3 - (1/2)$. The rule of thumb leads then to an error which is approximately
\begin{equation}
\sqrt{2\pi}C_1 (\lambda C_2)^{1/2-C_3}
{\rm e}^{-\frac{1}{C_2\lambda}}
\label{eq:minerr}
\end{equation}
This function is an approximate envelope for the accuracy curves at successive orders as 
illustrated in Fig. \ref{fig:anh}.

This type estimate is not always correct. For instance, for the ground state of the double-well 
potential, the instanton effect is much larger than this estimate. Often, neither the asymptotic behavior nor the non-perturbative effects are known. Consequently, it is worth 
trying to figure out a general method to handle the problem.

\section{Quantum Field Theory with a Large Field Cutoff}

As can be seen in several examples where a path integral formulation is available, the 
factorial growth of perturbative series is related to configurations with arbitrary large fields. 
Large order coefficients are dominated by large field configurations for which the expansion 
of the exponential of the perturbation at that order is not a good approximation. 
In two non-trivial examples \cite{convpert} where the modified coefficients 
can be calculated numerically, it was shown that a large field cutoff drastically 
affects the asymptotic behavior of the perturbative series. At this point we are only able 
to do numerical calculations for specific types of field cutoffs. Namely, we remove the tails of 
integration in configuration space. If $\phi_x$ denotes a scalar field at sites $x$ of a lattice, we impose the condition $|\phi_x |<\phi_{max}$ at every site. 
This procedure is quite convenient for a Monte Carlo calculation, but not to write 
modified Feynman rules because the additional condition is non-local in momentum space. 
At this point, a large field cutoff procedure that leads to simple 
Feynman rules remains to be found. 

Quantum field theory with a large field cutoff is a subject in infancy. 
Numerical studies \cite{asymptuniv,tractable} show that, after appropriate rescalings, the transition 
between the large field cutoff regime and the small field cutoff regime can be described 
in good approximation by a universal function. This hints at a renormalization group explanation. 
It might be possible to construct an effective action, with couplings  running with the field cutoff. 

\section{Optimization and Interpolation}

At a given order $K$ and coupling $\lambda$, we can adjust $\phi_{max}(\lambda,K)$ in order to minimize or eliminate the discrepancy with the (usually unknown) correct value.  As $\phi_{max}$ is varied, the curve (or the derivative) of the approximate expression crosses the numerical curve (or its derivative).
The strong coupling can be used to calculate approximately this optimal $\phi_{max}(\lambda,K)$ and so this is 
a natural approach for the interpolation between the weak and strong coupling regimes. 
This procedure has been illustrated with two examples\cite{optim,plaquette}.
The calculation of the modified coefficients $a_k(\phi_{max})$ fall in 
three categories. 
Low $k$ (the usual ones with exponentially small corrections calculable semi-classically);
intermediate $k$ (crossover; complicated but with universal features) and 
large $k$ (power suppressed; no $k!$ behavior). 
The intuitive flow picture is that 
the beginning of the series corresponds to the behavior of the scaling variables near 
the gaussian fixed point.
The large order, corresponds to the approach of the high-temperature/strong-coupling fixed point.
The coefficients in the crossover ($\phi_{max}$ dependent) correspond to the crossover in the flows.

\section{Lattice Gauge Theory with One Plaquette}
A simple example where the general ideas discussed above can be easily implemented is a $SU(N)$ lattice gauge theory with one plaquette \cite{plaquette}. The partition function reads.
\begin{equation}
Z(\beta,N)=\int \prod_{l\in p} dU_l {\rm e} ^{-\beta(1-\frac{1}{N}Re TrU_p)}\ ,
\end{equation}
with $\beta=2N/g^2$.
Specializing to $N=2$, fixing the gauge to the identity for three of the links and integrating over the angular variables, we obtain
\begin{equation}
	Z(\beta,2)=(2/\beta)^{3/2}\frac{1}{\pi}\int_0^{2\beta}dt t^{1/2}
	{\rm e}^{-t}\sqrt{1-(t/2\beta)}\ .
\label{eq:tint}
\end{equation}
Note that that due to the compactness of the group, the partition function has a 
large field cutoff, and also a large action cutoff which is gauge invariant.
To construct a weak coupling expansion, we expand the square root in power of $\beta ^{-1}$. 
However, the ``coefficients'' depend on $\beta$ because the range of integration does. 
To get a regular series, we add the tails of integration because they are ${\rm e}^{-2\beta}={\rm e}^{-8/g^2}$ effects, but this affects the asymptotic behavior of the series 
and the coefficients now grow factorially. The optimal order to truncate this series $2\beta 
-1/2$. Incidently, this is also the order where the peak of the integrand moves out of the 
range of integration if it is kept below $2\beta$ as in the exact expression Eq. (\ref{eq:tint}).

It is possible \cite{plaquette} 
to keep the finite bound of integration or to modify it in order to minimize the difference 
between the series at a given order and the original integral as explained in the previous section. However, if the order is large enough, the optimal cutoff becomes close to $2\beta$.
Note also that if we simply consider the regular perturbative series truncated using the rule of thumb, 
it is easy in this simple example to 
define and calculate the non-perturbative part of the integral. It consists in the 
higher order terms (to be calculated with the finite range of integration) minus the tails of integration that we have added. 

\section{The Non-Perturbative Part of the Plaquette}

In the previous example, we can obtain a converging expansion by calculating the 
coefficients of the $\beta^{-1}$ expansion keeping a finite range of integration. 
When there is more than one plaquette, this is much more complicated and we are presently 
developing practical methods to perform these calculations. In the rest of this talk, we discuss the regular perturbative series of the average plaquette of lattice QCD in the quenched approximation. 
We define the average plaquette
\begin{equation}
\label{eq:pdef}
P(\beta)\equiv (1/\mathcal{N}_p)\left\langle \sum_p
(1-(1/N)Re Tr(U_p))\right\rangle \ .
\end{equation}
with $\mathcal{N}_p$ the total number of plaquettes. The $\beta^{-1}$ series of $P$ has been calculated \cite{direnzo2000,rakow05} up to order 16.
Despite the long series available, the factorial growth is not apparent. 
Instead, the series appears more like it has finite radius of convergence \cite{rakow2002,third}. This is impossible 
because $P$ takes different limits \cite{gluodyn04} when $g^2\rightarrow 0^{\pm}$. The only plausible explanation is that the partition function has a zero in the complex $\beta$ plane near $\beta\simeq 5.7$. 
Zeroes have been found \cite{alan} for imaginary values of $\beta$ within predicted bounds \cite{third} (and not found below the lower bound), 
however their numerical significance remains to be established .

A simplified model \cite{mueller93,itep,burgio97} that is capable of producing 
a series with coefficients growing factorially is 
\begin{equation}
P \approx K \int_{t_1}^{t_2}dt {\rm e}^{-\bar{\beta}t}\ (1-t\ 33/16\pi^2)^{-1-204/121} \ .
\label{eq:mue}
\end{equation}
The new parameter $\bar{\beta}$ is related to the lattice $\beta$ by a relation of the form
$\bar{\beta}=\beta(1+d_1/\beta+\dots)$. When deriving this expression as a sum of bubble diagrams, one realizes that 
$t_1=0$ corresponds to momenta at the UV cutoff and $t_2=16\pi^2/33$ corresponds to the Landau pole. 

Note that the expression is quite similar to the one plaquette integral discussed in the previous section. This could in principle be compared with
what would be obtained from the probability distribution for one plaquette after integrating over all the other links. Note also that in his ITEP lectures, M. Shifman emphasizes that it is necessary to introduce the 
gluon condensate in order to keep $t_2$ low enough and regularize the perturbative 
series, exactly as we advocated in lattice gauge theory.

Expanding $(1-t\ 33/16\pi^2)^{-1-204/121}$ in powers of $t$ and extending the integration range to $\infty$, one finds at leading order that the 
coefficients of the $\bar{\beta}^{-1}$ expansion  grow like 
$(33/16\pi^2)^k\ \Gamma[k+204/121]$. 
According to the rule of thumb, we should truncate the series at an order $4.79\bar{\beta}-2.19$. For $\beta=6$ and $d_1$ small this means an order of about 25. 
For $d_1=-3$, this order is lowered to 12. 
Using Eq. (\ref{eq:minerr}) and shifting to the 
lattice $\beta$, we conclude that 
the error is proportional to 
\begin{equation}
(\beta)^{204/121-1/2}{\rm e}^{-(16\pi^2/33)\beta} \ .
\label{eq:guessintmodel}
\end{equation}
Except for the power -1/2, this expression depends on $\beta$ as the fourth power of the 
two-loop renormalization group invariant scale. In practice, the dependence on the power of 
$\beta$ is quite weak in the region of $\beta$ where an empirical envelope for the accuracy 
curves can be seen. In Fig. 2, the thick line has been drawn using the one-loop formula formula  with an adjustable normalization constant $1.3 \times 10^{10}\times {\rm e}^{-(16\pi^2/33)\beta}$ and looks like a decent envelope.
\begin{figure}
\begin{center}
\includegraphics[width=0.8\textwidth]{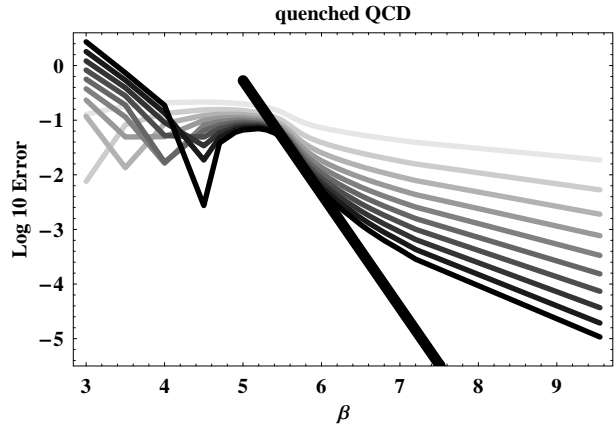}
\caption{Difference between the perturbative series at order 1 to 10 and the numerical value of the plaquette (in Log 10 scale). As the order increases, 
the curves get darker. The thick curve is discussed in the text. 
}
\end{center}
\end{figure}
When the order increases, will the accuracy curves reach the straight line or will they maintain some curvature? Higher order extrapolations \cite{rakow2002,rakow05} favor the 
second possibility with a power 4 of the force scale. Lower power of this scale provide 
good fits of the accuracy curves at various orders. This question will be discussed into more detail in a forthcoming publication\cite{preprint}. 
\section{Conclusions}
A better control of perturbative series of the standard model is necessary.
A field cutoff drastically improves the asymptotic behavior of series. In simple examples, 
the field cutoff can be chosen to minimize the discrepancy with the uncut theory.
There seems to be a connection between the crossover behavior of the perturbative coefficients 
and crossover behavior of the RG flows.
Numerical methods need to be developed to implement large field or large action cutoffs.
Analytic methods remain to be developed to estimate the various parameters determined empirically while studying the difference between the perturbative series and numerical values in lattice QCD.


\section{References}
\begin{thebibliography}{20}

\bibitem{dyson52}
F.~Dyson, {\em Phys. Rev.} {\bf 85}, 631 (1952).

\bibitem{zvibern}
Z.~Bern, A tour of the s-matrix: QCD, SYM and Supergravity, these Proceedings.

\bibitem{larin94}
S.~A. Larin, T.~van Ritbergen and J.~A.~M. Vermaseren, {\em Phys. Lett.} {\bf
  B320}, 159 (1994).

\bibitem{convpert}
Y.~Meurice, {\em Phys. Rev. Lett.} {\bf 88}, 141601 (2002).

\bibitem{asymptuniv}
L.~Li and Y.~Meurice, {\em J. Phys.} {\bf A39}, 8681 (2006).

\bibitem{tractable}
L.~Li and Y.~Meurice, {\em J. Phys. A} {\bf 38}, 8139 (2005).

\bibitem{optim}
B.~Kessler, L.~Li and Y.~Meurice, {\em Phys. Rev.} {\bf D69}, 045014 (2004).

\bibitem{plaquette}
L.~Li and Y.~Meurice, {\em Phys. Rev.} {\bf D71}, 054509 (2005).

\bibitem{direnzo2000}
F.~Di~Renzo and L.~Scorzato, {\em JHEP} {\bf 10}, 038 (2001).

\bibitem{rakow05}
P.~E.~L. Rakow, {\em PoS} {\bf LAT2005}, 284 (2006).



\bibitem{rakow2002}
R.~Horsley, P.~E.~L. Rakow and G.~Schierholz, {\em Nucl. Phys. Proc. Suppl.}
  {\bf 106}, 870 (2002).



\bibitem{third}
L.~Li and Y.~Meurice, {\em Phys. Rev.} {\bf D73}, 036006 (2006).

\bibitem{gluodyn04}
L.~Li and Y.~Meurice, {\em Phys. Rev. D} {\bf 71}, 016008 (2005).


\bibitem{alan}
A. Denbleyker and Y. Meurice, work in progress.

\bibitem{mueller93}
A.~H. Mueller Talk given at Workshop on QCD: 20 Years Later, Aachen, Germany,
  9-13 Jun 1992.

\bibitem{itep}
M.~Shifman, {\em ITEP Lectures on Particle Physics and Field Theory}, 2001).

\bibitem{burgio97}
G.~Burgio, F.~Di~Renzo, G.~Marchesini and E.~Onofri, {\em Phys. Lett.} {\bf
  B422}, 219 (1998).

\bibitem{preprint}
Y.Meurice, preprint in progress.

\end{thebibliography}

\end{document}